%% file: main2.tex
\documentclass{ifacconf}

\usepackage{graphicx,amsmath,amssymb}
\usepackage{comment,color}
\usepackage{natbib}   
\usepackage{tikz}
\usepackage{url}
\usepackage{pgfplots}
\usepackage{grffile}
\pgfplotsset{compat=newest}
\usepgfplotslibrary{groupplots}   
\usetikzlibrary{plotmarks}
\usetikzlibrary{arrows.meta}
\usepgfplotslibrary{patchplots}
\usepackage{siunitx}
\usetikzlibrary{shapes,arrows}
\tikzstyle{block} = [draw, fill=white, rectangle]
\tikzstyle{sum} = [draw, fill=white, circle, node distance=1cm]
\usetikzlibrary{decorations.pathmorphing}
\usetikzlibrary{calc}
\usetikzlibrary{arrows.meta,positioning}
\usetikzlibrary{calc}
\tikzset{
  block/.style = {draw, fill=white, rectangle, minimum height=2em, minimum width=4em, align=center},
  line/.style  = {-{Stealth[length=2.2mm]}, thick},
  plain/.style = {thick}
}
\usetikzlibrary{fit,backgrounds}

\usepackage{enumitem}

\newcommand{\bbR}{\ensuremath{{\mathbb R}}}
\newcommand{\bbZ}{\ensuremath{{\mathbb Z}}}
\newcommand{\bbN}{\ensuremath{{\mathbb N}}}

\newcommand{\calI}{\mathcal{I}}
\newcommand{\calP}{\mathcal{P}}

\newcommand{\sfs}{\mathsf{s}}
\newcommand{\sfL}{\mathsf{L}}

\newcommand{\bfr}{\mathbf{r}}
\newcommand{\bfc}{\mathbf{c}}

\newcommand{\bfz}{\mathbf{z}}
\newcommand{\bfzero}{\mathbf{0}}
\newcommand{\bfone}{\mathbf{1}}

\newcommand{\Enc}{\mathsf{Enc}}
\newcommand{\Dec}{\mathsf{Dec}}

\newcommand{\sk}{\mathsf{sk}}

\newcommand{\ini}{\mathsf{ini}}

\newcommand{\diag}{\mathrm{diag}}

\newcommand{\llceil}{\left\lceil}

\newcommand{\rrfloor}{\right\rfloor}
\newcommand{\modp}{~\mathrm{mod}~}

\begin{document}
\begin{frontmatter}

\title{Sensor Attack Detection Method for Encrypted State Observers}

\thanks[footnoteinfo]{This work was supported by the National Research Foundation of Korea (NRF) grant funded by the Korea government (MSIT) (No. RS-2022-00165417 and RS-2024-00353032).}

\author[First]{Yeongjun Jang} 
\author[Second]{Sangwon Lee} 
\author[Second]{Junsoo Kim}

\address[First]{ASRI, Department of Electrical and Computer Engineering, Seoul National University, Seoul, Korea (email: jangyj0512@snu.ac.kr)}
\address[Second]{Department of Electrical and Information Engineering, Seoul National University of Science and Technology, Seoul, Korea (email: leesangwon@cdslst.kr, junsookim@seoultech.ac.kr)}

\begin{abstract}                
This paper proposes an encrypted state observer that is capable of detecting sensor attacks without decryption. 
We first design a state observer that operates over a finite field of integers with the modular arithmetic.
The observer generates a residue signal that, under sparse attack and sensing redundancy conditions, indicates the presence of attacks.
Then, we develop a homomorphic encryption scheme that enables the observer to operate over encrypted data while automatically disclosing the residue signal.
Unlike our previous work restricted to single-input single-output systems, the proposed scheme is applicable to general multi-input multi-output systems.
Given that the disclosed residue signal remains below a prescribed threshold, the full state can be recovered as an encrypted message.
\end{abstract}


\end{frontmatter}

\section{Introduction}
Enhancing the security of networked control systems has attracted growing interest, as successful attacks can cause severe physical and/or economic damage.
In this context, encrypted control (\cite{KogiFuji15,KimjKimd22,SchlBinf23}) has emerged as a method to protect data in both the communication and computation layers with the use of homomorphic encryption---a cryptosystem that enables arithmetic operations to be evaluated directly over encrypted data without decryption.
By exploiting this property, control computations can be securely outsourced to untrusted remote servers without sharing the decryption key.

While encryption preserves data confidentiality, it simultaneously hides the effect of data corruption attacks.
Therefore, ensuring both confidentiality and integrity of data has been a critical yet challenging problem in encrypted control.
\cite{FausZhan24} proposed a resilient homomorphic encryption scheme that neutralizes the effect of additive attacks that lie within a certain range. 
Towards attack detection, \cite{MartZhan19,AlexBurb22} incorporated anomaly detectors into encrypted control systems, which trigger an alarm when a residue signal exceeds a prescribed threshold. 
However, because the residue signal is also encrypted, it needs to be sent to an external detector holding the decryption key, thus incurring additional communication burden.


In this paper, we propose an encrypted state observer that can directly detect sensor attacks without decryption.
Towards this end, we first design a state observer that operates over a finite field of integers with the modular arithmetic, as homomorphic encryption schemes are typically built upon such fields.
The observer generates a residue signal that, under sparse attack and sensing redundancy conditions, indicates the presence of attacks.
In particular, even though the observer's state and signals may ``overflow'' the modulus range due to potentially arbitrary and/or unbounded attacks, the proposed residue signal still enables attack detection.

\begin{figure*}[t]
\centering
\input{diagram}
\caption{ Configuration of the proposed encrypted state observer capable of detecting sensor attacks. 
}
\label{fig:diagram}
\end{figure*}

Then, we develop a Learning With Errors (LWE) (\cite{Rege09}) based homomorphic encryption scheme that enables the designed observer to operate over encrypted data while automatically disclosing the residue signal. 
The key idea is to modify the standard LWE based scheme by leveraging the zero-dynamics of the observer, so that the ``masking term'' of the encrypted residue, which conceals the underlying message, is enforced to remain identically zero.
This mechanism was first introduced in our previous work (\cite{JangLeej}) for single-input single-output (SISO) systems.
To accommodate our multi-input multi-output observer handling arbitrary sparse sensor attacks, as a conference version, we extend the scheme to the multi-input single-output (MISO) case and apply it repeatedly to each output channel, making it applicable to general multi-input multi-output systems.
The proposed scheme is secure in the sense that it does not compromise the security of the standard LWE based scheme beyond the intentional disclosure of the residue signal.

\textit{Notation:} 
The sets of integers, non-negative integers, positive integers, and real numbers are denoted by $\bbZ$, $\bbZ_{\ge 0}$, $\bbN$, and $\bbR$, respectively.
For $p\in\bbN$, we define $[p]:=\{1,2,\ldots,p\}$.
For real vectors (matrices), $\|\cdot\|$ denotes the (induced) infinity-norm. 
The identity and the zero matrix are denoted by $I$ and $\bfzero$, respectively, with their dimensions indicated as subscripts when necessary.
For a sequence $v_1,\ldots, v_n$ of scalars, vectors, or matrices, with an index set $\Lambda=\{\lambda_1<\cdots<\lambda_{|\Lambda|}\}\subset [n]$, we define $[v_1;\cdots;v_n]:=[v_1^\top,\ldots,v_n^\top]^\top$ and $v_\Lambda:=[v_{\lambda_1};\cdots;v_{\lambda_{|\Lambda|}}]$.

\section{Preliminaries and Problem Setting}

\subsection{LWE based Homomorphic Encryption Scheme}\label{subsec:LWE}
We briefly introduce the LWE based encryption scheme of \cite{Rege09}, focusing on its additively homomorphic property.
For $q\in\bbN$, we consider $\bbZ_q:=\bbZ\cap [-q/2,q/2)$ as the space of messages to be encrypted. 
The modulo operation that maps $\bbZ$ onto $\bbZ_q$ is defined as $a\modp q:=a - \lfloor (a+q/2)/q\rfloor q\in\bbZ_q$ for all $a\in\bbZ$, which applies component-wisely to vectors and matrices.
Given a secret key $\sk\in\bbZ_q^N$ of length $N\in\bbN$, an $h$-dimensional message $m\in\bbZ_q^h$ is encrypted as
\begin{align}\label{eq:enc}
    \Enc(m):=
    \begin{bmatrix}
        m + b &~ A    
    \end{bmatrix} \modp q \in \bbZ_q^{h\times(N+1)},    
\end{align} 
where $A\in\bbZ_q^{h\times N}$ is a randomly generated matrix, and $b:=A\cdot \sk + e \modp q\in\bbZ_q^h$ is the ``masking term'' that conceals the message. 
Here, $e\in\bbZ^h$ is a small ``error term'' bounded as $\|e\|\le \Delta$ for some $\Delta>0$.
The decryption of an $h$-dimensional ciphertext (encrypted message) $\bfc\in\bbZ_q^{h \times(N+1)}$ is performed as
\begin{equation*}
    \Dec(\bfc) := \bfc \begin{bmatrix}
        1 \\ -\sk
    \end{bmatrix} \modp q \in \bbZ_q^h,
\end{equation*}
so that the original message can be approximately recovered as $\Dec(\Enc(m))=m+e\modp q$.
For the sake of simplicity, we omit the modulo operation in the arguments of $\Enc$ and $\Dec$ throughout the paper.

The described scheme is \textit{additively homomorphic}, that is, 
\begin{equation}\label{eq:homoAdd}
    \Dec(\bfc_1 + \bfc_2) = \Dec(\bfc_1) + \Dec(\bfc_2)~\modp q
\end{equation}
holds for any $\bfc_1\in\bbZ_q^{h\times (N+1)}$ and $\bfc_2\in\bbZ_q^{h\times (N+1)}$.
From this property, it follows that for any integer matrix $K\in\bbZ^{d\times h}$,
\begin{align*}
    K\cdot \Enc(m) := 
    \begin{bmatrix}
        Km + Kb &~ KA    
    \end{bmatrix} \modp q \in \bbZ_q^{d \times (N+1)}
\end{align*} 
is a $d$-dimensional ciphertext, which can be decrypted as 
\begin{align}\label{eq:homoMult}
    \Dec(K\cdot \Enc(m)) &= Km+Ke\modp q. 
\end{align}
Note that $K$ is multiplied to the message $m$, as well as the error term $e$.
We refer to $Km$, $Kb$, and $Ke$ as the message, masking term, and error term of $K\cdot\Enc(m)$, respectively.

\subsection{Problem Setting}

Consider a discrete-time plant written by 
\begin{subequations}\label{eq:sys}
\begin{align}
    x(t+1) &= Ax(t) + Bu(t),~~~~x(0)=x_\ini,\label{eq:sysState} \\
    y(t) &= Cx(t) + a(t), \label{eq:sysOutput}
\end{align}
\end{subequations}
where $x(t)\in\bbR^n$ is the state with the initial value $x_\ini\in\bbR^n$, $u(t)\in\bbR^m$ is the control input, $y(t)\in\bbR^p$ is the sensor output, and $a(t)\in\bbR^p$ is the sensor attack signal.
In particular, $u(t)$ is assumed to be a known nominal input, possibly generated by a feedback controller independent of the proposed observer, so that we can focus on observer design.
We assume that $u(t)$ and $y(t)$ remain bounded for all $t\ge 0$ in the absence of attacks, i.e., when $a(t)\equiv \bfzero$.

Our objective is to design an encrypted state observer for \eqref{eq:sys} that receives encryptions of $u(t)$ and $y(t)$, and computes its next state and the state estimate over encrypted data using the homomorphic properties in \eqref{eq:homoAdd} and \eqref{eq:homoMult}.
The main challenge is that detecting sensor attacks becomes difficult as all signals remain encrypted. 
In this motivation, we suggest to modify the encryption algorithm $\Enc$, so that a residue signal, which indicates the presence of attacks, is automatically disclosed.
By utilizing the disclosed residue signal, the proposed encrypted observer can directly detect sensor attacks without requiring access to the secret key.

We conclude the section by assuming the sparsity of the attack signal and the redundant observability of \eqref{eq:sys}, as is common in the literature on resilient state observers (see \cite{KimjLeej18,LeecShim18} and references therein).  
For each $i\in[p]$, the $i$-th sensor output is denoted by
\begin{equation}\label{eq:output}
    y_i(t)=C_ix(t) + a_i(t)\in\bbR,
\end{equation}
where $C_i\in\bbR^{1\times n}$ is the $i$-th row of $C$ and $a_i(t)\in\bbR$ is the $i$-th component of $a(t)$.
We do not impose any restriction on $a(t)$; it may be arbitrary and/or unbounded.
Instead, we restrict the attacker's resources by assuming that at most $k$ sensors, $k<p$, can be compromised. 
Under this constraint, we further assume that the plant is $k$-redundant observable, meaning that it remains observable after removing any set of at most $k$ sensors.
This condition rules out undetectable attacks and is known to be equivalent to sparse attack detectability (\cite{LeecShim18}).
\begin{assum}\label{asm:attack}
    There exists an integer $k<p$ such that at least $p-k$ sensors are not compromised for all $t\ge 0$. 
    That is, the set $\calI:=\left\{i\in[p] \mid a_i(t)\equiv 0 \right\}$
    satisfies $|\calI|\ge p -k$.
\end{assum}

\begin{assum}\label{asm:obsv}
    For any subset $\Lambda\subset [p]$ such that $|\Lambda|\ge p-k$, the pair $(A,C_\Lambda)$ is observable.
\end{assum}

\section{Attack Detection over $\bbZ_q$}

In what follows, we construct a state observer defined over $\bbZ_q$, which will serve as the basis for the proposed encrypted observer.
For each $i\in[p]$, let $l_i\in\bbZ_{\ge 0}$ denote the observability index of the pair $(A,C_i)$. 
Using the Kalman observable decomposition (\cite{Chen}), the \textit{observable subsystem} of \eqref{eq:sysState} with \eqref{eq:output} can be written as
\begin{subequations}\label{eq:obsvSysFull}
\begin{align}\label{eq:obsvSys}
    z_i(t+1) &= F_i z_i(t) + \Phi_iB u(t),~~~z_i(0)=z_{i,\ini},  \\
    y_i(t) &= J_i z_i(t) + a_i(t), \label{eq:partialOutput} 
\end{align}
\end{subequations}
where $z_i(t)=\Phi_ix(t)\in\bbR^{l_i}$ is the observable substate with some full row rank matrix $\Phi_i\in\bbR^{l_i\times n}$.
Since the pair $(F_i,J_i)$ is observable by construction, we assume without loss of generality that \eqref{eq:obsvSysFull} is given in the observable canonical form, i.e., 
\begin{align*}
    F_i=
    \begin{bmatrix}
        0 & \cdots & 0 & f_{i,1} \\
        1 & \cdots & 0 & f_{i,2} \\
        \vdots & \ddots & \vdots & \vdots \\
        0 & \cdots & 1 & f_{i,l_i}
    \end{bmatrix}\in\bbR^{l_i\times l_i},~~~~
    J_i =
    \begin{bmatrix}
        0 & \cdots & 0 & 1    
    \end{bmatrix} \in \bbR^{1\times l_i}.
\end{align*}

A ``partial'' observer for $z_i(t)$ is constructed from $y_i(t)$, as
\begin{align}\label{eq:partObsv}
    \hat{z}_i(t+1) &= F_i \hat{z}_i(t) + \Phi_iBu(t) + L_i(y_i(t) - J_i \hat{z}_i(t))  \\
    &=: \bar{F_i}\hat{z}_i(t) + \Phi_iBu(t) + L_i y_i(t),~~\hat{z}_i(0)=\hat{z}_{i,\ini}, \nonumber
\end{align} 
where $\hat{z}_i(t)\in\bbR^{l_i}$ is the partial observer state with the initial value $\hat{z}_{i,\ini}\in\bbR^{l_i}$. 
In particular, we design the observer gain as $L_i=[f_{i,1};\cdots;f_{i,l_i}]\in\bbR^{l_i}$, so that the resulting state matrix 
\begin{align*}
    \bar{F}_i = F_i-L_iJ_i = 
    \begin{bmatrix}
        \bfzero & 0 \\
        I_{l_i-1} & \bfzero
    \end{bmatrix} \in \bbZ^{l_i\times l_i}
\end{align*}
is both Schur stable and integer-valued. 
The rationale is that the state matrix of a dynamic system needs to be an integer matrix to be encrypted, as shown in \cite{CheoHank18}.
By combining the partial observers \eqref{eq:partObsv} for all $i\in[p]$, the full observer is obtained as 
\begin{subequations}\label{eq:observer}
\begin{align}\label{eq:observerState}
    \hat{z}(t+1) = \bar{F}\hat{z}(t) + 
    \begin{bmatrix}
            \Phi B & L
    \end{bmatrix} \!
    \begin{bmatrix}
        u(t) \\ y(t)    
    \end{bmatrix}, ~~~ \hat{z}(0) = \hat{z}_\ini,
\end{align}
where  
\begin{align*}
    \hat{z}(t)\!&:=\!\begin{bmatrix}
        \hat{z}_1(t);\cdots;\hat{z}_p(t)
    \end{bmatrix}\!\in\!\bbR^{l},&\!\!\! \! 
    \hat{z}_\ini \!&:=\!\begin{bmatrix}
        \hat{z}_{1,\ini};\cdots;\hat{z}_{p,\ini}
    \end{bmatrix}\!\in\!\bbR^{l}, \nonumber \\
    \bar{F}\! &:=\! \diag(\bar{F}_1,\ldots,\bar{F}_p) \!\in\!\bbZ^{l\times l}, &\!\!\!\!
    \Phi\!&:=\!   
    \begin{bmatrix}
        \Phi_1;\cdots;\Phi_p
    \end{bmatrix}\!\in\!\bbR^{l\times n}, \nonumber \\
    L\!&:=\! \diag(L_1, \ldots, L_p) \!\in\! \bbR^{l\times p},&\!\!\!\!l\!&:=\!\textstyle\sum_{i\in[p]}l_i
\end{align*}
with $\diag(\cdot)$ denoting the block-diagonal matrix operator.

When $a(t)\equiv \bfzero$, $\hat{z}(t)$ reaches $z(t):=[z_1(t);\cdots;z_p(t)]$ in finite time due to the nilpotency of $\bar{F}$.
Moreover, since $\Phi$ has full column rank by Assumption~\ref{asm:obsv}, we have $x(t)=\Phi^\dagger\Phi x(t) = \Phi^\dagger z(t)$, where $(\cdot)^\dagger$ denotes the Moore-Penrose inverse.
Consequently, in this case, the state $x(t)$ can be exactly recovered in finite time as $\hat{x}(t):= \Phi^\dagger \hat{z}(t)$.

However, since some of the sensors may be corrupted, the estimate $\hat{x}(t)$ cannot be used directly.
To address this, we introduce a collection of index sets, as
\begin{align*}
    \calP:= \{\Lambda \subset [p] \mid |\Lambda| = p-k\}.
\end{align*}
By Assumption~\ref{asm:obsv}, $\Phi_\Lambda$ has full column rank for every $\Lambda\in\calP$, and hence $x(t) = \Phi_\Lambda^\dagger z_\Lambda(t)$. 
Furthermore, Assumption~\ref{asm:attack} ensures the existence of at least one uncorrupted index subset $\Lambda\in\calP$, i.e., $\Lambda \subset \calI$, for which $\hat{z}_\Lambda(t)$ reaches $z_\Lambda(t)$.
Since such $\Lambda$ cannot be identified a priori, we fix an ordering $(\Lambda_1,\ldots,\Lambda_{|\calP|})$ of $\calP$ and define the residue signal
\begin{align}\label{eq:resDef}
    \hat{r}(t) := 
    \begin{bmatrix}
        \hat{x}_{\Lambda_1}(t) - \hat{x}(t)   \\ 
        \vdots      \\
        \hat{x}_{\Lambda_{|\calP|}}(t) - \hat{x}(t) 
    \end{bmatrix} \in\bbR^{n_r},
\end{align}
\end{subequations}
where 
$\hat{x}_{\Lambda}(t) := \Phi_\Lambda^\dagger \hat{z}_\Lambda(t)\in\bbR^n$ for all $\Lambda\in\calP$ and 
$n_r:=n|\calP|$. 
Roughly, a small $\|\hat{r}(t)\|$ implies that $\hat{x}(t)$ remains close to the estimate obtained from an uncorrupted index subset, and can therefore serve as a reliable estimate.

\begin{rem}
    In \cite{KimjLeej18,LeecShim18}, the residual signal was defined as $\hat{r}(t) = (I-\Phi\Phi^\dagger)\hat{z}(t)$, which represents the deviation of $\hat{z}(t)$ from the image space of $\Phi$, where $z(t)$ resides.
    However, this geometric interpretation does not translate naturally to $\bbZ_q$ because the modulo operation may truncate the higher bits of the state and signals (especially when some components of $\hat z(t)$ are arbitrarily corrupted by attacks), thereby destroying the underlying geometric structure.
    This led to the definition of a new residue signal in \eqref{eq:resDef}.
\end{rem}

We now convert the observer \eqref{eq:observer} to operate over $\bbZ_q$.
First, we scale and round the matrices in \eqref{eq:observer}, except for the integer matrix $\bar{F}$, as
\begin{align}\label{eq:GHbar}
    \bar{G} &:=  \llceil \frac{\begin{bmatrix}
        \Phi B & L
    \end{bmatrix}}{\sfs_1} \rrfloor  \in \bbZ^{l \times (m+p)}, \\ \overline{\Phi_\Lambda^\dagger} &:= \llceil \frac{\Phi_\Lambda^\dagger}{\sfs_1} \rrfloor \in \bbZ^{n \times l_\Lambda },~~\forall \Lambda\in\calP, \nonumber 
\end{align} 
and $\overline{\Phi^\dagger}:=\lceil \Phi^\dagger/\sfs_1 \rfloor$,
where $1/\sfs_1\ge1$ is a scale factor and $l_\Lambda = \sum_{i\in\Lambda} l_i$.
Similarly, the initial value $\hat{z}_{\ini}$, and the input signals $u(t)$ and $y(t)$ 
of the observer are quantized as
\begin{align}\label{eq:IniInbar}
    \bar{z}_{\ini} &:= \llceil \frac{\hat{z}_{\ini}}{\sfs_1\sfs_2} \rrfloor \modp q \in \bbZ_q^l, \\
    \bar{v}(t) &:=  \llceil  \frac{\begin{bmatrix}
        u(t) ; y(t)
    \end{bmatrix}}{\sfs_2} \rrfloor \modp q \in \bbZ_q^{m+p}, \nonumber
\end{align}
using an additional scale factor $1/\sfs_2\ge1$.

As a result, the quantized observer over $\bbZ_q$ is obtained as
\begin{subequations}\label{eq:observerQuant}
\begin{align}
    \bar{z}(t+1) &= \bar{F}\bar{z}(t) + \bar{G} \bar{v}(t)\modp q,\\
    \bar{z}(0) &= \bar{z}_{\ini} \modp q, \nonumber
\end{align}
where $\bar{z}(t)=[\bar{z}_1(t);\cdots;\bar{z}_p(t)]\in\bbZ_q^{l}$ is the quantized observer state with $\bar{z}_i(t)\in\bbZ_q^{l_i}$ for each $i\in[p]$. 
The quantized residue signal $\bar{r}(t)\in\bbZ_q^{n_r}$ can be computed as 
\begin{align}\label{eq:observerQuantResidue}
    \bar{r}(t)
    =  \begin{bmatrix}
        \bar{x}_{\Lambda_1}(t) - \bar{x}(t)\\
        \vdots \\
        \bar{x}_{\Lambda_{|\calP|}}(t) - \bar{x}(t) 
    \end{bmatrix} 
     \modp q =: \bar{H}\bar{z}(t) \modp q
\end{align}
\end{subequations}
with some appropriate matrix $\bar{H}\in\bbZ^{n_r \times l}$,
where
$ \bar{x}_\Lambda(t) :=\overline{\Phi_\Lambda^\dagger}\bar{z}_\Lambda(t) \modp q \in\bbZ_q^n$ for all $\Lambda\in\calP$, and $\bar{x}(t):=\overline{\Phi^\dagger}\bar{z}(t)\modp q\in\bbZ_q^n$.

The following theorem states that sensor attacks can be detected by monitoring whether a suitably scaled $\bar{r}(t)$ exceeds a prescribed threshold, provided that $q$ is chosen sufficiently large. 
To state the result, we define 
\begin{align*}
    \tilde{z}_\ini \!:=\! \max_{i\in[p]} \left\|z_{i,\ini} \!-\! \hat{z}_{i,\ini} \right\|, ~
    \kappa \!:=\! \max \left\{\! \left\|\Phi^\dagger \right\|, \max_{\Lambda \in \calP} \left\{ \left\| \Phi_\Lambda^\dagger \right\| \right\} \!\right\}.
\end{align*}
Also, we define an indicator function $\bfone_{\{t< l_{\max}\}}$ that equals $1$ when $t<l_{\max}:=\max_{i\in[p]} l_i$, and $0$ otherwise.
In addition, the stability of $\bar{F}$ and the boundedness of the signals of \eqref{eq:sys} imply that there exists $M>0$ such that
\begin{align}\label{eq:boundM}
    \sup_{t \ge 0}\left\{ \left\| \hat{r}(t)\right\|, \, \|\hat{z}(t)\| \right\} \le M,
\end{align}
when $\calI=[p]$, i.e., $a(t)\equiv \bfzero$.

\begin{thm}\label{thm:obsverQuant}
    For any $\epsilon>0$, there exist $\sfs_1'>0$ and $\sfs_2'>0$ such that for any $\sfs_1<\sfs_1'$, $\sfs_2<\sfs_2'$, and 
    \begin{align}\label{eq:qbound}
        q >  2\frac{\kappa (M+ 2 \tilde{z}_\ini) + 2\epsilon}{\sfs_1^2\sfs_2}, 
    \end{align}
    the followings hold:
    \begin{enumerate}[leftmargin=*]
        \item If the inequality 
        \begin{align}\label{eq:criterion}
            \|\sfs_1^2\sfs_2 \cdot \bar{r}(t)\| &\le 2 \kappa \tilde{z}_\ini \cdot \bfone_{\{t< l_{\max}\}} + \epsilon
        \end{align}
        is violated for some $t\ge0$ then $\calI\ne[p]$, i.e., $a(t)\not\equiv \bfzero$.
        \item Under Assumptions~\ref{asm:attack} and~\ref{asm:obsv}, if \eqref{eq:criterion} holds then
        \begin{align}\label{eq:criterion2}
             \left\| x(t) - \sfs_1^2\sfs_2 \cdot \bar{x}(t) \right\| \le 3\kappa \tilde{z}_\ini \cdot \bfone_{\{t< l_{\max}\}} + 2\epsilon. 
        \end{align}
    \end{enumerate}
\end{thm}
\begin{pf}
    The proof is omitted due to space limitations and can be found in \cite{JangLees25}.
\end{pf}

Theorem~\ref{thm:obsverQuant} establishes that the violation of \eqref{eq:criterion} indicates the presence of a sensor attack.
Conversely, the satisfaction of \eqref{eq:criterion} does not guarantee the absence of an attack.
Nonetheless, it ensures that the effect of any existing attack on the state estimate remains sufficiently small, so that the state $x(t)$ can be recovered in the sense of \eqref{eq:criterion2} with a bounded error.
Choosing the scale factors $\sfs_1$ and $\sfs_2$ sufficiently small serves to reduce the precision losses caused by the rounding operations in \eqref{eq:GHbar} and \eqref{eq:IniInbar}.
Moreover, the condition \eqref{eq:qbound} ensures that the modulus $q$ is large enough to prevent overflow, i.e., loss of higher bits of $\bar{z}(t)$ due to the modulo operation, when attack-free.

\section{Encrypted Attack Detection}

This section presents the proposed encrypted state observer capable of detecting sensor attacks.
We develop a modified LWE based encryption scheme that enables the observer \eqref{eq:observerQuant} to be implemented over encrypted data, while selectively disclosing the residue signal $\bar{r}(t)$.
First, we analyze the zero-dynamics of MISO systems over $\bbZ_q$.

\subsection{Zero-dynamics of MISO Systems over $\bbZ_q$}\label{subsec:normal}

Let us fix an index $j\in[n_r]$ and consider the following MISO system over $\bbZ_q$:
\begin{align}\label{eq:bDyn}
    b_z(t+1) &= \bar{F}b_z(t) + \bar{G} b_v(t)\modp q,~~~b_z(0)=b_\ini, \nonumber \\
    b_r(t) &= \bar{H}^{(j)}b_z(t)\modp q, 
\end{align}
where $\bar{H}^{(j)}$ is the $j$-th row of $\bar{H}$, $b_z(t)\in\bbZ_q^l$ is the state with the initial value $b_\ini\in\bbZ_q^l$, $b_v(t)\in\bbZ_q^{m+p}$ is the input, and $b_r(t)\in\bbZ_q$ is the output.

Throughout this section, we choose $q$ to be prime, so that $\bbZ_q$ becomes a field.
This allows us to use standard linear algebraic notions (e.g., rank, linear independence, matrix inversion), which have been developed for arbitrary fields in \cite[Chapter~1]{Frie14}.
Consequently, we can define the relative degree of \eqref{eq:bDyn}, analogously to \cite{Khal96}, as the smallest integer $\nu\ge 1$ that satisfies
\begin{align}\label{eq:rankCond}
        \bar{H}^{(j)}\bar{F}^h\bar{G}\modp q&=\bfzero,~~~ \forall h=0,1,\ldots,\nu-2, \\
        \bar{H}^{(j)}\bar{F}^{\nu-1}\bar{G}\modp q &\ne \bfzero. \nonumber
    \end{align}
Although $\nu$ depends on the index $j$, we omit its dependency for notational simplicity; the same convention applies to all scalars and matrices in this subsection.

The following proposition presents the normal form representation of \eqref{eq:bDyn}. 

\begin{prop}\label{prop:normal}
    Suppose that the system \eqref{eq:bDyn} has relative degree $\nu \ge 1$.
    Then, there exists an invertible matrix $[T_1;T_2]\in\bbZ_q^{l \times l}$ such that the coordinate transformation
    \begin{align*}
        \begin{bmatrix}
            b_\xi(t) \\ b_w(t)
        \end{bmatrix} := \begin{bmatrix}
            T_1 \\ T_2
        \end{bmatrix} b_z(t) \modp q,
    \end{align*} 
    with $b_\xi(t)\in\bbZ_q^{l-\nu}$ and $b_w(t)\in\bbZ_q^\nu$,
    yields the \textit{normal form} of \eqref{eq:bDyn} written by
    \begin{align}\label{eq:normal}
        b_\xi(t+1) \!&=\! S_1 b_\xi(t) + S_2 b_w(t) + S_3 b_v(t) \modp q, \\
        b_{w_1} (t+1) \!&=\! b_{w_2}(t), \nonumber \\
        &\vdots  \nonumber \\
        b_{w_{\nu-1}} (t+1) &= b_{w_{\nu}} (t), \nonumber \\
        b_{w_{\nu}} (t+1) &= \Psi b_\xi(t) + \Gamma b_w(t) + \Sigma b_v(t)\modp q
        ,  \nonumber \\ 
        b_r(t) &= b_{w_1}(t), \nonumber
    \end{align}
    for some matrices $S_1\in \bbZ^{(l-\nu) \times (l-\nu)}$, $S_2\in \bbZ^{(l-\nu) \times \nu}$, $S_3\in \bbZ^{(l-\nu) \times (m+p)}$, $\Psi\in\bbZ^{1\times (l-\nu)}$, $\Gamma\in\bbZ^{1\times \nu}$, and $\Sigma\in\bbZ^{1\times (m+p)}$,
    where $b_{w}(t) =:[b_{w_1}(t);\cdots;b_{w_{\nu}}(t)]$.
\end{prop}
\begin{pf}
    The proof and explicit expressions for the associated matrices are omitted due to space limitations and can be found in \cite{JangLees25}.
\end{pf}

The obtained normal form provides a clear interpretation of the necessary and sufficient condition under which the output $b_r(t)$ remains identically zero. 
The proof can be found in \cite{JangLees25}.

\begin{lem}\label{lem:zeroingCond}
    Suppose that the system \eqref{eq:bDyn} has relative degree $\nu\ge 1$. 
    Then, $b_r(t)\equiv 0$ if and only if
    \begin{subequations}\label{eq:zeroingCond}
    \begin{align}\label{eq:zeroingCondIni}
        b_w(0) &=\bfzero, \\
        b_v(t) &= - \Sigma^\dagger\Psi b_\xi(t)+(I-\Sigma^\dagger\Sigma)b_\mu(t) \modp q, \label{eq:zeroingCondInput}
    \end{align}
    \end{subequations}
    for some $b_\mu(t)\in\bbZ_q^{m+p}$.
\end{lem}

Since $\Sigma$ is a non-zero row vector, it holds that
$\Sigma \Sigma^\dagger=1$. 
Hence, $b_w(t)\equiv \bfzero$ under \eqref{eq:zeroingCond}, and
the dynamics of $b_\xi(t)$ is given by 
\begin{align}\label{eq:zeroDyn}
    b_\xi(t+1)  = S b_\xi(t) + S_3(I-\Sigma^\dagger \Sigma)b_\mu(t) \modp q,
\end{align}
where $S:=S_1 - S_3 \Sigma^\dagger\Psi$.
We refer to \eqref{eq:zeroDyn} as the \textit{zero-dynamics} of the system \eqref{eq:bDyn} parameterized by the signal $b_\mu(\cdot)$. 
It describes the internal dynamics of \eqref{eq:normal} consistent with the constraint $b_r(t)\equiv 0$ under \eqref{eq:zeroingCond}.

By exploiting the zero-dynamics, we can explicitly compute the portions of the initial condition $b_\ini$ and the input signal $b_v(\cdot)$ that, when canceled out, ensure $b_r(t)\equiv 0$.
To emphasize the dependence of $b_r(t)$ on the initial condition and the input sequence, we often write
\begin{align*}
    b_r(t) = b_r(t\mid b_\ini, b_v(\cdot)).
\end{align*}
To state the following proposition, let $[V_1,V_2] := [T_1;T_2]^{-1}$ with $V_2\in\bbZ_q^{l \times \nu}$, so that $T_2V_1=\bfzero$ and $T_2V_2=I_\nu$.

\begin{prop}\label{prop:cancel}
    Suppose that the system \eqref{eq:bDyn} has relative degree $\nu \ge 1$.
    Given $b_\ini\in\bbZ_q^l$ and $b_v(\cdot): \bbZ_{\ge 0} \to \bbZ_q^{m+p}$, there exist
    $\tilde{b}_\ini\in\bbZ_q^\nu$ and $\tilde{b}_v(\cdot): \bbZ_{\ge 0} \to \bbZ_q$ such that
    \begin{align}\label{eq:cancelToShow}
        b_r(t\mid b_\ini - V_2\tilde{b}_\ini, b_v(\cdot)-\Sigma^\dagger \tilde{b}_v(\cdot)) \equiv 0,
    \end{align}
    which are uniquely determined by 
    \begin{align}
        \tilde{b}_\ini &= T_2b_\ini \modp q, \label{eq:bw'}\\
        \tilde{b}_v(t) &= \Sigma b_v(t) + \Psi b_\xi(t) \modp q,~~~\forall t\ge0, \label{eq:bv'}
    \end{align}
    where $b_\xi(t)$ is the solution to \eqref{eq:zeroDyn} with $b_\xi(0)=T_1b_\ini\modp q$ and $b_\mu(t)\equiv b_v(t)$.
\end{prop}
\begin{pf}
    The proof is omitted due to space limitations and can be found in \cite{JangLees25}.
\end{pf}

\subsection{Proposed Encryption Scheme and Encrypted Observer}\label{subsec:encObsv}

We now describe the proposed encryption scheme and the construction of the corresponding encrypted observer.
Let the initial condition and input of \eqref{eq:observerQuant} be scaled by a scale factor $\sfL\in\bbN$ and be encrypted as
\begin{align}\label{eq:standardEnc}
    \Enc(\sfL\cdot\bar{z}_\ini) &= 
    \begin{bmatrix}
        \sfL\cdot\bar{z}_\ini + b_\ini &~ A_\ini    
    \end{bmatrix} \modp q, \\ 
    \Enc(\sfL\cdot\bar{v}(t)) &= 
    \begin{bmatrix}
        \sfL\cdot\bar{v}(t) + b_v(t) &~ A_v(t)    
    \end{bmatrix} \modp q, \nonumber
\end{align}
where $A_\ini \in \bbZ_q^{l\times N}$ and $A_v(t)\in\bbZ_q^{(m+p)\times N}$ are randomly generated matrices, and $b_\ini\in\bbZ_q^l$ and $b_v(t)\in\bbZ_q^{m+p}$ are the corresponding masking terms defined as in \eqref{eq:enc}, respectively.
The scale factor $\sfL$ is introduced to negate the effect of the error terms injected during encryption.

For each $j\in[n_r]$, following the observation of Proposition~\ref{prop:cancel}, we modify \eqref{eq:standardEnc} and define the encryption algorithms $\Enc_\ini^{(j)}(\cdot): \bbZ_q^l \to \bbZ_q^{l\times(N+2)}$ and $\Enc_t^{(j)}(\cdot): \bbZ_q^{m+p} \to \bbZ_q^{(m+p)\times(N+2)}$ for all $t\ge 0$, as
\begin{align}\label{eq:encModified}
    &\Enc_\ini^{(j)}(\sfL\cdot\bar{z}_\ini)\!:=\!
    \begin{bmatrix}
        \sfL\cdot\bar{z}_\ini+ b_\ini - V_2\tilde{b}_\ini, &A_\ini, & V_2\tilde{b}_\ini
    \end{bmatrix}\modp q, \nonumber \\
    &\Enc_t^{(j)}(\sfL\cdot\bar{v}(t)) \\
    &\!:=\! \begin{bmatrix}
        \sfL\cdot\bar{v}(t)+b_v(t)-\Sigma^\dagger \tilde{b}_v(t), & A_v(t), &\Sigma^\dagger \tilde{b}_v(t)
    \end{bmatrix}\modp q, \nonumber
\end{align}
where $\{\tilde{b}_\ini,\tilde{b}_v(t),V_2,\Sigma\}$ are computed as in Section~\ref{subsec:normal} with respect to the index $j$.
Since the ciphertexts in \eqref{eq:encModified} have one additional column compared to those in \eqref{eq:standardEnc}, we define the decryption of a ciphertext $\bfc\in\bbZ_q^{h \times (N+2)}$ as
\begin{align*}
    \Dec'(\bfc) := \bfc \begin{bmatrix}
        1 \\ -\sk \\ 1
    \end{bmatrix}  \modp q \in \bbZ_q^h.
\end{align*}
Then, it can be easily verified that the modified encryption scheme is also additively homomorphic.

With the proposed encryption scheme, we construct an encrypted state observer for each $j\in[n_r]$, as
\begin{subequations}\label{eq:encObsv}
\begin{align}\label{eq:encObsvState}
    \bfz^{(j)}(t+1) &= 
    \bar{F} \cdot \bfz^{(j)}(t) + 
    \bar{G}\cdot \Enc_t^{(j)}\left( \sfL\cdot \bar{v}(t)\right)
    \modp q, \nonumber \\
    \bfz^{(j)}(0) &= \Enc_\ini^{(j)}\left( \sfL\cdot \bar{z}_\ini \right), 
\end{align}
where $\bfz^{(j)}(t)\in \bbZ_q^{l\times (N+2)}$ is the encrypted state.
The encrypted residue signal is then computed as 
\begin{align}
    \bfr(t) := 
    \begin{bmatrix}
        \bar{H}^{(1)}\cdot \bfz^{(1)}(t)\\ \vdots \\ \bar{H}^{(n_r)} \cdot\bfz^{(n_r)}(t)    
    \end{bmatrix} \modp q\in \bbZ_q^{n_r \times(N+2)}.
\end{align} 
\end{subequations}
The full configuration of the proposed encrypted state observer is illustrated in Fig.~\ref{fig:diagram}.

The following theorem states that the residue signal $\bar{r}(t)$ of \eqref{eq:observerQuant} can be exactly recovered without decryption by appropriately scaling the first column of $\bfr(t)$. 
In addition, $\bar{x}(t)$ can also be exactly recovered by decrypting and post-processing the encrypted state for any $j\in[n_r]$, provided that the scale factors $\{\sfL,\sfs_1,\sfs_2\}$ and the modulus $q$ are chosen appropriately.

\begin{thm}\label{thm:disclose}
    Consider the observer \eqref{eq:observerQuant} over $\bbZ_q$ and the corresponding encrypted observer \eqref{eq:encObsv}. 
    \begin{enumerate}
        \item Let $\bfr_1(t)\in \bbZ_q^{n_r}$ denote the first column of $\bfr(t)$.
        Then,
        \begin{align}\label{eq:rdisclose}
            \bfr_1(t) = \sfL\cdot \bar{r}(t)\modp q
        \end{align}
        for all $t\ge 0$.
        \item Under Assumptions~\ref{asm:attack} and~\ref{asm:obsv}, for any $\epsilon>0$, there exist $\sfs_1'>0$ and $\sfs_2'>0$ such that for any $\sfs_1<\sfs_1'$, $\sfs_2<\sfs_2'$, and
        \begin{subequations}
        \begin{align}
            \sfL &> 2\left(\frac{\kappa}{\sfs_1} +\frac{l}{2}\right)\left(1+l_{\max}\left\|\bar{G}\right\| \right)\Delta, \label{eq:Lbound}\\
            q &> \sfL\left(2\frac{\kappa (M+ 2 \tilde{z}_\ini) + 2\epsilon}{\sfs_1^2\sfs_2} + \frac{1}{2} \right), \label{eq:qbound2}
        \end{align}
        \end{subequations}
        if \eqref{eq:criterion} holds then  
        \begin{align}\label{eq:stateRecover}
            \!\!\llceil \!\frac{\overline{\Phi^\dagger}\Dec'(\bfz^{(j)}(t))\modp q}{\sfL} \!\rrfloor  \!=\! \overline{\Phi^\dagger}\bar{z}(t)\modp q \!=\! \bar{x}(t)
        \end{align}
        for all $j\in[n_r]$.
    \end{enumerate}
\end{thm}
\begin{pf}
    The proof is omitted due to space limitations and can be found in \cite{JangLees25}.
\end{pf}
In particular, \eqref{eq:rdisclose} implies that $\bar{r}(t)$ can be recovered from $\bfr_1(t)$ by multiplying the multiplicative inverse of $\sfL$ in $\bbZ_q$.
As a result, Theorem~\ref{thm:disclose} ensures that the proposed encrypted observer can directly detect sensor attacks without decryption, while recovering the state as a ciphertext.

One might be concerned that disclosing the residue signal $\bar{r}(t)$ could leak some sensitive or private information.
However, it only reflects the differences $\bar{x}_{\Lambda_i}(t)-\bar{x}(t)$ and does not directly reveal the plant state itself. 
Moreover, residue signals are typically dominated by noise, disturbances, and model uncertainties in practice, so it will be difficult to recover meaningful information solely from $\bar{r}(t)$; see \cite[Remark~4]{JangLeej} for related discussion. 

Finally, it is emphasized that our modification does not compromise the security of the standard LWE based scheme beyond the intentional disclosure of the residue signal. 
This follows from the same argument as in \cite[Theorem~2]{JangLeej} that the modified ciphertexts can be constructed from the standard ciphertexts and the residue signal, and vice versa.

\begin{rem}
    The proposed method encrypts the input signal $\bar{v}(t)$ separately for each $j\in[n_r]$ and runs $n_r$ MISO encrypted observers. 
    This can be massively parallelized, for example, using graphics processing units (GPUs).
    The computational burden could be further reduced by extending the analysis in Section~\ref{subsec:normal} to multi-input multi-output systems over $\bbZ_q$, which we leave for future work.
\end{rem}

\section{Simulation Results}

This section provides simulation results\footnote{ The code is fully available at \url{https://github.com/CDSL-EncryptedControl/enc_atk_detect}} of the proposed method applied to the three-inertia system of \cite{Ogat95}, using the same parameters as in \cite{LeecShim18}.
A model of the form \eqref{eq:sys} is obtained by discretizing the system with a sampling time of $\SI{0.1}{\s}$.
From this model, we obtain $n=6$, $m=1$, $p=5$, $l=24$, and $l_{\max}=6$.
Since our focus lies in constructing a state observer, we employed a simple state feedback nominal controller of the form $u(t)=Kx(t)$ 
that renders $A+BK$ Schur stable.
Explicit expressions for $A$, $B$, $C$, and $K$ are provided in \cite{JangLees25}.

The encryption parameters are chosen as $(N,\Delta,q)=(2^{12},19.2,2^{109}-31)$ to ensure $128$-bit security (\cite{AlbrChas21}), where $q$ is a prime.
We set $\epsilon = 0.3$ and chose the scale factors as $\sfL=2^{44}$ and $\sfs_1=\sfs_2=10^{-5}$ according to Theorems~\ref{thm:obsverQuant} and~\ref{thm:disclose}.
The initial conditions of the plant \eqref{eq:sys} and the observer \eqref{eq:observer} are chosen as $x_\ini = [1;1;1;1;1;1]$ and $\hat{z}_\ini=0$.

Note that the pair $(A,C)$ is $2$-redundant observable, meaning that Assumption~\ref{asm:obsv} holds with $k=2$.
Accordingly, we applied the attack signal illustrated in Fig.~\ref{fig:result}--(a) to the third sensor at $t=\SI{2.5}{\s}$. 
In practice, an adversary would compromise the ciphertexts in \eqref{eq:encModified} by injecting integer-valued attacks in the first column in which the message resides. 
For simplicity, we equivalently modeled this as an additive attack applied directly to $y(t)$.

Fig.~\ref{fig:result}--(b) and Fig.~\ref{fig:result}--(c) show the state estimation error and the residue signal obtained from the encrypted observer \eqref{eq:encObsv}, after recovering $\bar{r}(t)$ and $\bar{x}(t)$ via \eqref{eq:rdisclose} and \eqref{eq:stateRecover}, respectively.
As shown in the figures, both the estimation error and the residue signal exceed the threshold specified in Theorem~\ref{thm:obsverQuant} in the presence of attacks, confirming that attacks can be directly detected without decryption.

\begin{figure}[t]
    \begin{centering}
        \input{img/myplot}
    \end{centering}
    \vspace{-7mm}
    \caption{Simulation results. (a) Injected sensor attack signal $a(t)$. (b--c) State estimation error and residue signal obtained from the encrypted observer \eqref{eq:encObsv} (blue solid) and the thresholds in Theorem~\ref{thm:obsverQuant} (red dashed).}
    \label{fig:result}
\end{figure}

\section{Conclusion}
We have proposed an encrypted state observer capable of detecting sensor attacks.
By exploiting the zero-dynamics of MISO systems over the field $\bbZ_q$, we developed a modified LWE based encryption scheme that automatically discloses the residue signal. 
As a result, the proposed method enables anomaly detection without access to the secret key, while recovering the full state as an encrypted message.
Future work will focus on disclosing only a binary alarm signal indicating whether the residue signal exceeds a prescribed threshold, while also encrypting the control parameters.

\bibliography{ifacconf}

\end{document}

%% file: diagram.tex
\begin{tikzpicture}[font=\small]

\tikzset{
  sum/.style={
    circle,
    draw,
    minimum size=5mm,
    inner sep=0pt
  }
}


\node[block] (plant) {Plant};

\node[sum, right=0.5cm of plant] (sumY) {\scriptsize $+$};

\node[block, right=0.5cm of sumY] (sensor) {Sensor};

\node[block, below=0.8cm of sensor] (stackQ)
{Quantize \\ {\scriptsize\eqref{eq:IniInbar}}};

\coordinate[left=1cm of plant] (uIn);
\coordinate[above=0.8cm of sumY] (aIn);

\coordinate[left=0.5cm of plant] (uBranch);
\coordinate[below=0.95cm of plant] (uBelow);
\coordinate (uRouteRight) at ($(stackQ.west)+(0,0)$);


\draw[line]
    (uIn) -- node[midway, above] {$u(t)$} (plant.west);

\draw[line]
    (sumY) -- (sensor.west);

\draw[line]
    (uBranch) |- (uRouteRight);

\draw[line]
    (plant.east) -- (sumY.west);

\draw[line]
    (aIn) -- node[pos=0.2, right] {$a(t)$} (sumY.north);

\draw[line]
    (sensor.south) -- node[midway, right] {$y(t)$} (stackQ.north);

\node[block, right=1.2cm of stackQ] (EncLast)
{$\Enc_t^{(n_r)}$};
\draw[line]
    (stackQ.east) --  (EncLast.west);

\node[block, above=1cm of EncLast] (EncFirst)
{$\Enc_t^{(1)}$};

\coordinate (branchpt) at ($(stackQ.east)!0.7!(EncLast.west)$);

\draw[line] (branchpt) |- node[midway, right, yshift = -9mm] {$\bar v(t)$} (EncFirst.west);

\node at ($(EncFirst)!0.45!(EncLast)$) {\bf$\vdots$};

\node[block, right=0.8cm of EncFirst] (obsvFirst)
{Encrypted \\ Observer {\scriptsize\eqref{eq:encObsvState}}
};

\draw[line] (EncFirst) -- (obsvFirst);

\node[block, right=0.8cm of EncLast] (obsvLast)
{Encrypted \\ Observer {\scriptsize\eqref{eq:encObsvState}}
};

\draw[line] (EncLast) -- (obsvLast);

\node at ($(obsvFirst)!0.45!(obsvLast)$) {\bf$\vdots$};

\coordinate (obsvMid) at ($(obsvFirst)!0.5!(obsvLast)$);

\node[block, anchor=west] (scaling) at ([xshift=2.5cm]obsvMid)
{Scaling \\ 
{\scriptsize\eqref{eq:rdisclose}}};

\coordinate (merge) at ([xshift=-12mm]scaling.west);
\draw[plain] (obsvFirst.east) -| (merge);
\draw[plain] (obsvLast.east)  -| (merge);
\draw[line] (merge) --node[midway, above] {$\bfr_1(t)$} (scaling.west);

\node[block, right=0.8cm of scaling] (detector)
{Anomaly \\ Detector {\scriptsize\eqref{eq:criterion}}};
\draw[line] (scaling) --node[midway, above] {$\bar{r}(t)$} (detector);

\begin{pgfonlayer}{background}

\node[
    draw,
    dotted,
    thick,
    rounded corners,
    fit=(uIn)(plant)(sumY)(stackQ)(uBelow)(uRouteRight)(branchpt)(EncFirst)(EncLast),
    inner sep=6pt
] (boxPlant) {};

\node[anchor=south west]
    at ($(boxPlant.north west)+(0pt,1pt)$)
    {\bf Plant side};

\node[
    draw,
    dotted,
    thick,
    rounded corners,
    fit=(obsvFirst)(obsvLast)(merge)(scaling)(detector),
    inner sep=6pt
] (boxNetwork) {};

\node[anchor=south west]
    at ($(boxNetwork.north west)+(0pt,1pt)$)
    {\bf Network side};

\end{pgfonlayer}

\end{tikzpicture}

%% file: img/myplot.tex
\begin{tikzpicture}

\definecolor{darkgray176}{RGB}{176,176,176}
\definecolor{steelblue31119180}{RGB}{31,119,180}
\definecolor{myyellow}{rgb}{0.92900,0.69400,0.12500}%

\begin{groupplot}[group style={group size=1 by 3}]

\nextgroupplot[
scaled x ticks=manual:{}{\pgfmathparse{#1}},
tick align=outside,
tick pos=left,
title={(a)},
title style={
    at={(0.5,-0.15)},    
    anchor=north         
},
width=0.49\textwidth,
height=0.17\textwidth,
x grid style={darkgray176},
xmajorgrids,
xmin=0, xmax=49,
xtick style={color=black},
xticklabels={},
xtick={0,24,49},
y grid style={darkgray176},
ylabel={$a(t)$},
ylabel style={yshift=-6pt},
ymajorgrids,
scaled y ticks=false,
yticklabel style={/pgf/number format/fixed},
ymin=-1.1, ymax=1.1,
ytick={-1,0,1},
ytick style={color=black}
]
\addplot [line width=1.2pt, steelblue31119180]
table {%
0 0
1 0
2 0
3 0
4 0
5 0
6 0
7 0
8 0
9 0
10 0
11 0
12 0
13 0
14 0
15 0
16 0
17 0
18 0
19 0
20 0
21 0
22 0
23 0
24 0
25 1
26 1
27 1
28 1
29 1
30 0
31 0
32 0
33 0
34 0
35 0
36 0
37 0
38 0
39 0
40 0
41 0
42 0
43 -1
44 -1
45 0
46 0
47 0
48 0
49 0
};

\nextgroupplot[
scaled x ticks=manual:{}{\pgfmathparse{#1}},
tick align=outside,
tick pos=left,
width=0.49\textwidth,
height=0.17\textwidth,
title={(b)},
title style={
    at={(0.5,-0.15)},    
    anchor=north         
},
x grid style={darkgray176},
xmajorgrids,
xmin=0, xmax=49,
xtick style={color=black},
xticklabels={},
xtick={0,24,49},
y grid style={darkgray176},
ylabel={$\|x(t) - \sfs_1^2\sfs_2\cdot\bar{x}(t)\|$},
ymajorgrids,
ymin=-0.2, 
ymax=120,
ytick={0,120},
ytick style={color=black},
]
\addplot [line width=1.2pt, steelblue31119180]
table {%
0 0.984310235475424
1 2.43333438576514
2 2.72984801267664
3 2.17426844696356
4 1.61940831590993
5 2.50889401520737e-06
6 3.25167641712643e-06
7 6.08799620771538e-06
8 3.14577418442097e-06
9 3.97609378011143e-06
10 2.96468642557723e-06
11 5.55626556442146e-06
12 6.26588310898013e-06
13 5.23720211248468e-06
14 8.41978343035787e-06
15 5.25754553137858e-06
16 5.20641168647029e-06
17 3.18607145030625e-06
18 3.97189031081657e-06
19 3.61129934084436e-06
20 2.38923478426689e-06
21 1.46373231146857e-06
22 4.95828758215189e-06
23 9.93999113219068e-06
24 2.49362085789906e-06
25 4.40173106362292
26 1.84456763216797
27 3.5441505658865
28 2.84155660278047
29 2.10290568560094
30 1.68488857573041
31 0.872273206537614
32 1.96013581594147
33 1.21742182069499
34 1.28763867814023
35 4.83186778477834e-06
36 3.2565910039889e-06
37 6.28931736141106e-06
38 4.35435599281664e-06
39 2.79194279708198e-06
40 6.00683961537313e-06
41 4.99656631876289e-06
42 6.30076196006871e-06
43 4.40173280921857
44 1.84456535011382
45 2.10956787424729
46 0.996991159626193
47 1.44124302852781
48 1.21742055407344
49 1.28763745711479
};
\addplot [line width=1pt, red, dashed, const plot]
table {
    0   120
    6   0.6
    50  0.6
};

\coordinate (mainBLeft)  at (rel axis cs:0.49,0); 
\coordinate (mainBRight) at (rel axis cs:1,0);     

\nextgroupplot[
name = main,
width=0.49\textwidth,
height=0.17\textwidth,
tick align=outside,
tick pos=left,
title={(c)},
title style={
    at={(0.5,-0.8)},    
    anchor=north         
},
x grid style={darkgray176},
xlabel={Time ($\mathrm{sec}$)},
xlabel style={yshift=3pt},
xmajorgrids,
xmin=0, xmax=50,
xtick style={color=black},
y grid style={darkgray176},
ylabel={$\|\sfs_1^2\sfs_2\cdot\bar{r}(t)\|$},
ymajorgrids,
ymin=-0.2, ymax=81,
ytick={0, 80},
ytick style={color=black},
xtick={0,25,50},
xticklabels={0, 2.5, 5},
]
\addplot [line width=1.2pt,  steelblue31119180]
table {%
0 0
1 5.37294861330957
2 4.26127092155409
3 3.88139691035314
4 3.7303501344564
5 2.34411475390339
6 1.0800914884856e-05
7 5.884579855994e-06
8 7.34810734565e-06
9 5.357769409816e-06
10 6.268321422308e-06
11 4.410707663296e-06
12 6.008764779946e-06
13 5.907249310174e-06
14 9.293851657248e-06
15 1.2964214449739e-05
16 8.227734141345e-06
17 1.3457616534241e-05
18 7.766503644267e-06
19 6.245030055364e-06
20 6.019980775724e-06
21 6.492434516878e-06
22 9.325662481755e-06
23 9.293657618421e-06
24 8.747758325161e-06
25 5.907628166449e-06
26 5.38284222350954
27 2.8547001040613
28 6.64961754859499
29 2.8415575489428
30 5.27828688297608
31 2.66599089280808
32 2.33636966306167
33 3.93276546930452
34 2.07727560768808
35 2.56143807277914
36 3.199880101643e-06
37 4.575970212512e-06
38 6.029827588337e-06
39 5.828603837889e-06
40 8.273782958727e-06
41 4.938031141636e-06
42 1.0192557128088e-05
43 8.620250678241e-06
44 5.38284512297404
45 2.85470558774041
46 2.10957487090926
47 1.95492947678411
48 1.84247590011165
49 2.07727446297646
50 1.87183348594646
};
\addplot [line width=1pt, red, dashed, const plot]
table {
    0   80
    6   0.3
    50  0.3
};

\coordinate (mainLeft)  at (rel axis cs:0.5,0); 
\coordinate (mainRight) at (rel axis cs:1,0);   
\end{groupplot}

\begin{axis}[
   at={(main.north east)},     
   anchor=north east,
   xshift=-0.2cm, yshift=2.3cm,
   width=0.3\textwidth,
   height=0.14\textwidth,
   xmin=25, xmax=49,            
   ymin=-0.2, ymax=5,           
   ytick={0,5},
   ytick style={color=black},
   xtick={25,37,49},
   xticklabels={},              
   ticklabel style={font=\scriptsize},
   xmajorgrids,
   ymajorgrids,
]
\addplot [line width=1.2pt, steelblue31119180]
table {%
25 4.40173106362292
26 1.84456763216797
27 3.5441505658865
28 2.84155660278047
29 2.10290568560094
30 1.68488857573041
31 0.872273206537614
32 1.96013581594147
33 1.21742182069499
34 1.28763867814023
35 4.83186778477834e-06
36 3.2565910039889e-06
37 6.28931736141106e-06
38 4.35435599281664e-06
39 2.79194279708198e-06
40 6.00683961537313e-06
41 4.99656631876289e-06
42 6.30076196006871e-06
43 4.40173280921857
44 1.84456535011382
45 2.10956787424729
46 0.996991159626193
47 1.44124302852781
48 1.21742055407344
49 1.28763745711479
};

\addplot [line width=1pt, red, dashed, const plot]
table {
    25   0.6
    49   0.6
};

\coordinate (zoomBLeft)  at (rel axis cs:0,0);
\coordinate (zoomBRight) at (rel axis cs:1,0);
\end{axis}

\begin{axis}[
   at={(main.north east)},        
   anchor=north east,             
   xshift=-0.2cm, yshift=-0.2cm,  
   width=0.3\textwidth,
   height=0.14\textwidth,
   xmin=25, xmax=50,
   ymin=-0.2, ymax=8,    
   ytick={0, 4, 8},
   ytick style={color=black},
   xtick={25,37,50},
   xticklabels={},
   ticklabel style={font=\scriptsize},
   xmajorgrids,
   ymajorgrids,
]
\addplot [line width=1.2pt,  steelblue31119180]
table {%
25 5.907628166449e-06
26 5.38284222350954
27 2.8547001040613
28 6.64961754859499
29 2.8415575489428
30 5.27828688297608
31 2.66599089280808
32 2.33636966306167
33 3.93276546930452
34 2.07727560768808
35 2.56143807277914
36 3.199880101643e-06
37 4.575970212512e-06
38 6.029827588337e-06
39 5.828603837889e-06
40 8.273782958727e-06
41 4.938031141636e-06
42 1.0192557128088e-05
43 8.620250678241e-06
44 5.38284512297404
45 2.85470558774041
46 2.10957487090926
47 1.95492947678411
48 1.84247590011165
49 2.07727446297646
50 1.87183348594646
};
\addplot [line width=1pt, red, dashed, const plot]
table {
    20   0.3
    50  0.3
};
\coordinate (zoomLeft)  at (rel axis cs:0,0);
\coordinate (zoomRight) at (rel axis cs:1,0);
\end{axis}

\begin{scope}
  \draw[line width=1pt, dashed, gray] (mainLeft)  -- (zoomLeft);
  \draw[line width=1pt, dashed, gray] (mainRight) -- (zoomRight);
  
  \draw[line width=1pt, dashed, gray] (mainBLeft)  -- (zoomBLeft);
  \draw[line width=1pt, dashed, gray] (mainBRight) -- (zoomBRight);
\end{scope}

\end{tikzpicture}